\documentclass[pra,twocolumn,superscriptaddress,longbibliography,amsfonts,amssymb,amsmath,floatfix,floats,a4paper]{revtex4-1}

\usepackage [margin=1in]{geometry}
\usepackage{physics}
\usepackage[dvipdfmx]{graphicx}
\usepackage{mediabb}
\usepackage{amsfonts}
\usepackage{amsbsy}
\usepackage{latexsym}
\usepackage{mathrsfs}
\usepackage{bbm}
\usepackage{bm}
\usepackage{physics}
\usepackage{graphicx}

\begin{document}

\title{When photons are lying about where they have been}

\author{Lev Vaidman}
\affiliation{Raymond and Beverly Sackler School of Physics and Astronomy, Tel-Aviv University, Tel-Aviv 69978, Israel}

\author{Izumi Tsutsui}
\affiliation{Theory Center, Institute of Particle and Nuclear Studies, High Energy Accelerator Research Organization (KEK), Tsukuba 305-0801, Japan}

\noindent

\begin{abstract}
The past of the photon in a nested Mach-Zehnder interferometer with an inserted Dove prism is analyzed. It is argued that the Dove prism does not change the past of the photon. Alonso and Jordan  correctly point out  that an experiment by Danan et al. demonstrating the past of the photon in nested interferometer will show different results when the Dove prism is inserted. The reason, however, is not that the past is changed, but that the experimental demonstration becomes incorrect. The explanation of a signal from the place in which the photon was (almost) not present is given. Bohmian trajectory of the photon is specified.
\end{abstract}

\maketitle

\section{Introduction}

This work describes peculiar behaviour of photons in the modification of the experiment of Danan et al. \cite{Danan} proposed by Alonso and Jordan (AJ)
\cite{Jordan}. In the Danan et al. experiment photons were asked where exactly they have been inside a  nested interferometer tuned in a particular way.  The AJ
modification makes photons to tell that they have been in a place in which, according to the narrative of the two-state vector formalism (TSVF) \cite{past}
they could not have been. Note that this work is only slightly related to the results presented by one of the authors (L.V.)  at ``Emergent Quantum Mechanics''
which have been already published  \cite{Italy, VaidWein}.

\begin{figure}
\begin{center}
\includegraphics[width=7cm]{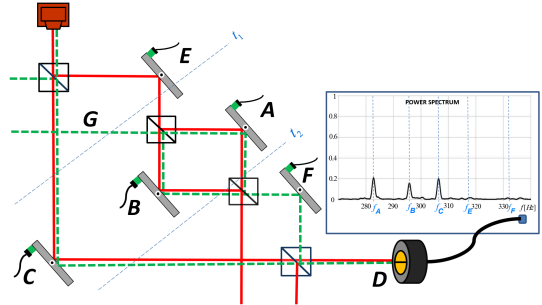}
\caption{Nested Mach-Zehnder interferometer with inner interferometer  tuned to destructive interference towards $F$.
Although our \lq common sense\rq\ suggests that the only possible path for the photon detected in $D$ is path $C$, the trace was found also inside the inner interferometer supporting the TSVF proposal according to which the particle was present in the places where forward (red continuous line) and backward (green dashed line) evolving wavefunctions overlap. The latter is demonstrated by the results of the measurement by Danan {\it et al}. \cite{Danan}.}
\label{fig:FIG-A}
\end{center}
\end{figure}

Textbooks of quantum mechanics teach us that we are not supposed to ask where
were the photons passing through an interferometer.  Wheeler \cite{Whe} introduced
the delayed choice experiment in an attempt to analyze this question.
Vaidman \cite{past} suggested a different approach.  He proposed a definition according to which a quantum particle was where it left a trace and showed that the past of the particle  can be easily seen in the framework of the TSVF \cite{AV90} as regions of the overlap of the forward and backward evolving quantum states. Vaidman, together with his collaborators, performed an experiment demonstrating a surprising trace of the photons in nested interferometers \cite{Danan}, see Fig.\ref{fig:FIG-A}. These results became a topic of a very large controversy \cite{LiCom, pastReply, pastsec, Saldaha, china_experiment, improved, PF, comment_potocek,Salih,comment_salih,
Bart,ComBart,RepComBart,Bula,Hashimi,Com-hashimi,Rep-Com-hashimi,Wu,Grif,Grif-com,Grif-rep,Sven,Sven-Com,Sven-Repl,SOK,SOK-REP,Nik,Nik-Com,
Nik-Resp,Dup,Sok-Dup,R-Sok-Dup,Zhou,Berge,BergeRep,Lima,Eli,Dissap,Hasegava}.

\section{Alonso and Jordan modified interferometer}

\begin{figure}
\begin{center}
\includegraphics[width=7cm]{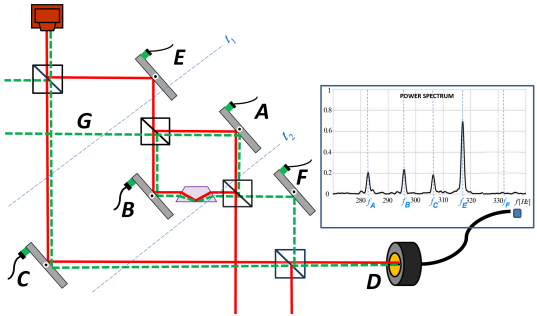}
\caption{Nested Mach-Zehnder interferometer with a Dove prism inside the inner interferometer as suggested by Alonso and Jordan \cite{Jordan}. The region of the overlap of the forward and the backward evolving states remains the same, but predicted results of an experiment similar to \cite{Danan} include a signal from mirror $E$ where the photon was not supposed to be.}
\label{fig:FIG-2}
\end{center}
\end{figure}

Here we  analyze, in our view, the most interesting objection which was made by Alonso and Jordan \cite{Jordan}. They suggested to insert a Dove prism inside one of the arms of  the inner interferometer (see Fig.\ref{fig:FIG-2}). They asked: \lq\lq Can a Dove prism change the past of a single photon?\rq\rq.
Their analysis of this modified experiment was correct. Although the  formalism suggested that the past of the photon remains the same as in the original experiment, {\it i.e.}, the photon was present near mirrors $C,  A,   B$ but not near mirrors $E$ and $F$, the experiment should show, in addition to frequencies  $f_C$, $f_A$, $f_B$, also the frequency $f_E$. This is  in contradiction with  the fact that the photons, according to Vaidman, were not present near mirror $E$.

The experiment of Danan {\it et al}. was not a direct measurement of the trace left by the photons. The reason is that such direct measurement is very difficult, as it requires collecting data about the trace conditioned on detection of the photon by a particular detector.
In the actual experiment, instead of measuring the trace on the external system (as in a recent experiment \cite{Steinberg}), the trace was \lq written\rq\ on the photons themselves, on the degree of freedom of their transverse motion.
Observing this degree of freedom of post-selected particle replaced the coincidence counting in the experimental setup.  Although indirect, the experiment \cite{Danan} was correct.
A local trace created at mirrors was read later on the quad-cell detector. We argue that introducing a Dove prism \cite{Jordan} spoils the experiment, making the signal at the quad-cell detector no longer a faithful representation of the trace created at mirror $E$.

Although the signal in the Danan {\it et al}. experiment was appearing as a particular frequency in the output of the quad-cell detector, the frequency was not an actual trace written on each photon.  Wiggling with different frequencies was a trick which allowed  in a single run to see records made at different mirrors. (It also improved significantly the signal to noise ratio, since noise had no preference to the frequencies of the wiggling mirrors.) The  physical signal in the Danan {\it et al}.  experiment (Fig.\ref{fig:FIG-A}) originated from the shift of the beam direction at a mirror. It  corresponded to  the transversal kick in the momentum $\delta_{p_x}$. This momentum shift translated into position shift of the beam which was read in the quad-cell detector. 
The property which allowed to observe the trace was  that  the change $\delta_{p_x}$ in the transversal momentum had no change when the beam  evolved toward port $D$ from all mirrors and through all possible paths. 

This is no longer the case when the Dove prism is introduced (Fig.\ref{fig:FIG-2}).
For mirrors $A$ and $C$, it is still true, since the modes do not pass through the Dove prism. 
For mirror $B$, 
there is a difference in that the Dove prism flips the sign of the signal. 
However, since we measure  just the size of the signal, this change is not observable, and the peak at frequency $f_B$ correctly signifies the presence of the photon in $B$.
The only problem occurs with the mirror $E$. The beam from $E$ reaches the detector through $A$ and through $B$. The shifts are in opposite direction, so reading position of the beam on the detector does not tells us what was the shift of the transversal momentum in $E$.
Therefore, we should not rely on the result of the experiment with the setup of the Danan {\it et al}. experiment when the Dove prism is present.

Note that a simple modification will restore the results of the Danan {\it et al}. experiment even with the presence of the Dove prism. If the wiggling of mirrors is made such that the beam is shifted in the direction perpendicular to the plane of the interferometer, the Dove prism will not cause flipping of the direction of the shift and the peak at $f_E$ will disappear.

\section{The trace analysis}

We have explained that the AJ modification of the Danan et al. experiment is not a legitimate experiment for measuring the presence (according the local trace definition)  of the particle near mirror $E$. Still it is of interest to understand how a strong signal with frequency $f_E$ is obtained in this modification. For this we need a more detailed analysis of traces  in the nested MZI experiments.  

We consider, for simplicity, an experiment in which only one particular mirror changes its angle at every run. The shift of the beam direction created at the mirror, characterized by the transversal momentum kick   $\delta_{p_x}$, leads to the shift of the beam position on the detector. This creates the signal: the difference in the current of the upper and the lower cells of the detector.

Let $\chi_0$ be the original mode of the photons without shifts.
The photons in a shifted beam will then be in a superposition of the original mode $\chi_0$ and a mode $\chi_\perp$ orthogonal to $\chi_0$:
\begin{equation}
|{\chi '}\rangle = \frac{1}{\sqrt{1+\epsilon^2}}\left(|{\chi_0}\rangle  + \epsilon |{\chi_\perp}\rangle\right).
\end{equation}

For small signals which appeared in the Danan  {\it et al}. experiment, the shift is proportional to the relative amplitude $\epsilon$ of the orthogonal mode \cite{universal}:
\begin{align}\label{eq::shiftOsingle}
 \delta p_x = 2 \epsilon \mathrm{Re} \left[ \langle \chi_0 |p_x| \chi_\perp \rangle \right] + \mathcal{O}(\epsilon^2).
\end{align}
Note that for a Gaussian beam (which is a good approximation of the beam in the experiment), higher order contributions do not appear \cite{universal}.

What is important for our analysis is that $\chi_0$ is symmetric with respect to the center of the beam in the transverse direction, while $\chi_\perp$, which can be approximated as a difference between two slightly shifted Gaussians, is an antisymmetric mode. Indeed, in momentum representation, we have
 \begin{equation}
\chi_0 \simeq {\cal N}_0\, e^{-\frac{p_x^2+p_y^2}{2\Delta^2}}, ~~~~~~ \chi_\perp \simeq {\cal{N}_\perp}\, p_x e^{-\frac{p_x^2+p_y^2}{2\Delta^2}} ,
\end{equation}
where $\Delta$ is the momentum uncertainty of the Gaussian beam, and ${\cal N}_0$, ${\cal N}_\perp$ are the normalization constants.

In the Danan {\it et al}. experiment (Fig.\ref{fig:FIG-A}), the trace of the photon was read as the shift of the beam on the detector. This shift is proportional to the strength of the trace quantified by the value of the relative amplitude $\epsilon$ of the orthogonal component. The original mode $\chi_0$ and the orthogonal mode $\chi_\perp$ evolve toward port $D$ from all mirrors and through all possible paths in an identical manner, so the position shift on the detector faithfully represents locally created trace.

This is no longer the case when the Dove prism is introduced (Fig.\ref{fig:FIG-2}). For mirror $B$, there is a difference: mode $\chi_0$ is unaffected by the presence of the prism, while mode $\chi_\perp$ flips the sign. The shift on the detector changes its direction. This change, however, is not observable in the experiment, since  the frequency spectrum is sensitive only to the size of the signal.
The observable difference appears for the mirror $E$.
There are two paths from $E$ to the output port $D$, one passing through mirror $A$ and another passing through mirror $B$.
The original symmetric mode $\chi_0$ would reach $D$ undisturbed both on path $A$ and on path $B$, while the orthogonal mode $\chi_\perp$ would reach $D$ undisturbed on path $A$ but with flipped sign on path $B$. When combined, however, there exists a phase difference $\pi$ between path $A$ and path $B$ which leads to destructive interference of the original symmetric mode and constructive interference of the orthogonal antisymmetric mode at the output port toward mirror $F$.   As a result, only the mode $\chi_\perp$ reaches $D$.

If we send the photon only in $A$ and do not move the mirror $A$, only mode $\chi_0$ reaches the detector. Small rotation of mirror $A$ will lead to appearance of mode $\chi_\perp$ with relative amplitude $\epsilon$. If, instead, in an undisturbed interferometer we send the photon only in channel $E$, nothing will reach the detector.  Small rotation of mirror $E$ will lead to appearance of mode $\chi_\perp$ on the detector and only mode  $\chi_\perp$. This mode by itself does not lead to a shift of the center of the beam on the detector. In the experiment, the photon is in a superposition of path $C$ and $E$. From path $C$ we get  mode $\chi_0$ with the same amplitude as it comes from path $A$. Interference between mode $\chi_0$, coming through $C$, and mode  $\chi_\perp$, coming through $E$, yields the shift  on the detector. And it is larger than the shift created by the same rotation of mirror $A$ because the intensity in $E$  is twice the intensity in $A$, and thus we have larger orthogonal component $\chi_\perp$ and therefore larger signal at $f_E$.

\section{Do the photons have any presence in E?}

Our analysis above shows that the experiment with the Dove prism does not contradict Vaidman's proposal \cite{past} demonstrated in the Danan {\it et al}. experiment, and explains using the standard quantum mechanics the appearance of the signal at frequency $f_E$. Thus, it provides a satisfactory reply to Alonso and Jordan.
However, it will also be of interest to explain the predicted results of Danan's setup with the Dove prism using Vaidman's approach.

Let us quote the Danan {\it et al.}~Letter \cite{Danan}:

\begin{quote}
``The photons themselves
tell us where they have been. And the story they tell is
surprising. The photons do not always follow continuous
trajectories. Some of them have been inside the nested
interferometer (otherwise they could not have known the
frequencies $f_A, \, f_B$), but they never entered and never left
the nested interferometer, since otherwise they could not
avoid the imprints of frequencies $f_E$ and $f_F$ of mirrors $E$
and $F$ leading photons into and out of the interferometer.''
\end{quote}

With the Dove prism present, however, we do get frequency $f_E$.
How can it happen if the photons were not in $E$ as we argued here? Let us analyse the situation, in which only mirror $E$ changes its angle by a small amount leading to the superposition (1) of the modes of the photon.

We start by repeating the analysis of the setup without the Dove prism in the framework of the TSVF \cite{AV90}.
After passing the mirror $E$, at time $t_1$, the forward evolving state is (see Fig.\ref{fig:FIG-2})
\begin{equation}\label{t_1forward}
|{\Psi}\rangle_{t_1}=\sqrt{\frac{{2}}{{3(1+\epsilon^2)}}}|{E}\rangle \left(|{\chi_0}\rangle+\epsilon |{\chi_\perp}\rangle \right)  + \frac{1}{\sqrt{3}}|{C}\rangle|{\chi_0}\rangle,
\end{equation}
where we split the which path and the mode degrees of freedom of the photon.
 The forward evolving state, at time $t_2$, in the middle of the interferometer is then
\begin{eqnarray}
\label{forward-t2}
|{\Psi}\rangle_{t_2}=&&\frac{1}{\sqrt{3(1+\epsilon^2)}} \left(|{A}\rangle + i|{B}\rangle\right)\left(|{\chi_0}\rangle +\epsilon |{\chi_\perp}\rangle \right)\nonumber\\ 
&&+\frac{1}{\sqrt{3}} |{C}\rangle|{\chi_0}\rangle.
\end{eqnarray}

Since in the experiment we use photon degrees of freedom for the measurement, we do not postselect on a particular state but rather on a space of states corresponding to all modes reaching detector $D$. So, strictly speaking, there is no definite backward evolving state. However, we can use a standard \lq trick\rq\ \cite{VaidWein}, in which we consider a hypothetical additional verification measurement of the mode state after  the postselection on the path $D$.  We verify that the state which we calculate will surely be there, and this verification measurement, together with the path post-selection, defines the backward evolving state.

The wave packets from $A$ and $B$ destructively interfere toward $F$ even when mirror $E$ is slightly rotated, so the only mode reaching $D$ is coming from $C$, which is $\chi_0$. Therefore, the backward evolving state starts from $\langle D|\langle {\chi_0}|$, which in the middle of the interferometer turns into
\begin{equation}
\langle{\Phi}|_{t_2}= \frac{1}{\sqrt{3}} \left(\langle{A}| - i\langle{B}|  + \langle{C}|\right)\langle{\chi_0}|.
\end{equation}
There is here destructive interference of the backward evolving quantum state toward $E$, so at time $t_1$, the backward evolving state is
\begin{equation}
\langle{\Phi}|_{t_1}=\frac{(\sqrt{2}\langle{G}| + \langle{C}|)\langle{\chi_0}|}{\sqrt{3}}.
\end{equation}
Thus, the weak value of the projection operator ${\rm \bf P}_E = |{E}\rangle\langle{E}|$ at $E$ is
\begin{equation}
({\rm \bf P}_E)_w=\frac{\langle{\Phi}| {\rm \bf P}_E |{\Psi}\rangle_{t_1}} { \langle{\Phi}|\Psi \rangle_{t_1}}=0 .
\end{equation}
Therefore, at time $t_1$   the photons have no presence in $E$, not even a ``small'' presence.

With the Dove prism inside, this is no longer the case. Instead of (\ref{forward-t2}) we obtain
\begin{eqnarray}
\label{forward-t2+dove}
|{\Psi'}\rangle_{t_2}=
&&\frac{1}{\sqrt{3(1+\epsilon^2)}}\big[(|{A}\rangle + i|{B}\rangle)|{\chi_0}\rangle \nonumber\\
&&+\epsilon (|{A}\rangle - i|{B}\rangle)|{\chi_\perp}\rangle )\big] + \frac{1}{\sqrt{3}} |{C}\rangle|{\chi_0}\rangle.
\end{eqnarray}
The wave packets from $A$ and $B$ destructively interfere toward $F$ for mode $\chi_0$, while the mode $\chi_\perp$ interfere constructively towards $F$. As a result, the backward evolving state (given the proper hypothetical measurement) starts approximately as
\begin{equation}
\frac{1}{\sqrt{1+2\epsilon^2}}\langle{D}|(\langle{\chi_0}|+\sqrt{2}\epsilon \langle{\chi_\perp}|).\end{equation}
Evolving it backwards until time $t_1$ we obtain approximately:
\begin{eqnarray}
\langle{\Phi'}|_{t_1}=
&&\frac{1}{\sqrt{3(1+2\epsilon^2)}}\big[(\sqrt{2}\langle{G}| + \langle{C}|)\langle{\chi_0}|  \nonumber\\
&&+\sqrt{2}\epsilon (\langle{C}| + \sqrt{2}\langle{E}|)\langle{\chi_\perp}|\big].
\end{eqnarray}

The Dove prism does not change the forward evolving state at $t_1$, so  even with the Dove prism, the state  is still given by (\ref{t_1forward}). Calculating now the  weak value of projection on $E$ yields
\begin{equation}\label{WV-Edove}
({\rm \bf P}_E)_w=\frac{\langle{\Phi'}| {\rm \bf P}_E |{\Psi}\rangle_{t_1}} { \langle{\Phi'}|\Psi \rangle_{t_1}}\simeq 2 \sqrt{2} \epsilon^2.
\end{equation}
The photon in the experiment with the Dove prism and the tilted mirror $E$ does have some presence in $E$. So, there is no clear paradox in obtaining the frequency $f_E$ which was present only in $E$ in the framework of the TSVF. 

One might wonder why there is no signal at $f_F$ similar to that  at $f_E$ in spite of the apparent symmetry of the experiment in the time symmetric TSVF. More careful analysis shows that the symmetry is not complete. When the mirror $F$ is tilted instead of mirror $E$,  inserting the Dove prism spoils the destructive interference of the backward evolving wave function towards $E$ similarly to spoiling interference toward $F$ by tilting mirror $E$. But titling mirror $E$ also changes the effective backward evolving state, while tilting mirror $F$ does not change the forward evolving  state. See details in the next section.

\section{Quantifying the presence of  photons }

The explanation of the peak at the frequency $f_E$ which we wish to provide is that the photon has a small presence there but the experimental records imprinted on the pre- and postselected photon reaching the detector are strong, so the size of the peak is similar to that of frequencies $f_C$, $f_B$, and $f_A$, where the photon presence is strong but the record is weak. However, the second order in $\epsilon$ for the presence of the photon in $E$ looks too small for this to be the case.  In more detail, 
for mirrors $A$, $B$, and $C$ the presence of the photon is of order 1 while the strength of the record is of size $\epsilon$. For mirror $E$, on the other hand, the presence characterized by  the weak value of projection operator (\ref{WV-Edove}) is apparently only of size $\epsilon^2$. The size of the record of an interaction is characterized by the created relative amplitude of the orthogonal component, see \cite{universal}. In our case, the record  created at $E$ which reaches the detector $D$ is represented by the orthogonal component $|{\chi_\perp}\rangle$ and it is the only component reaching the detector, since the symmetric component $|{\chi_0}\rangle$ is \lq filtered out\rq\ by the inner interferometer. 
Thus, we can say that the size of the record created at $E$ which reaches the detector is of order 1.   This naive consideration tells us that the peak at $f_E$ should be of order $\epsilon^2$ while other peaks are of order $\epsilon$, in contradiction with predicted results of the experiment which show that the peaks are of the same order.

It is true that the weak effects which depend only on the presence of the photon in $E$, such as the momentum transferred to the mirror $E$ when the photon bounces on the mirror, are proportional to $({\rm \bf P}_E)_w$, but the presence of a particle is defined according to {\it all} local traces it leaves, see Sec.VI of \cite{past}.
In our case, 
the weak value of the projection operator ${\rm \bf P}_E$ is not the correct parameter to quantify the presence of the particle. It is so, when the pre- and post-selection is on spatial degrees of freedom only, see \cite{universal}. Here, however, due to the postselection on a subspace, effectively, we are required to consider an associated postselection on a particular mode, along with the well defined preselected mode.
 Let us define an operator $O$ which connects between the mode $|{\chi_0}\rangle$ and the mode $|{\chi_\perp}\rangle$, possessing the eigenvalues $\pm 1$ for the states $|{\pm}\rangle = \left(|{\chi_0}\rangle\pm |{\chi_\perp}\rangle\right)/\sqrt {2}$.
For the experiment without the Dove prism, the weak value of local variable $O{\rm \bf P}_E$ still vanishes, but when the Dove prism is present, we have
\begin{equation}\label{WV-Edove2}
(O{\rm \bf P}_E)_w=\frac{\langle{\Phi'}| O{\rm \bf P}_E |{\Psi}\rangle_{t_1}} { \langle{\Phi'}|\Psi \rangle_{t_1}}\simeq 2 \sqrt 2 \epsilon.
\end{equation}
Therefore, the presence in $E$ is found to be of the order $\epsilon$ rather than $\epsilon^2$ which is obtained when we naively quantify the presence by $({\rm \bf P}_E)_w$. This explains why we obtain the signal from mirror $E$ of the same order as from other  mirrors. 

The weak value of local operator of order $\epsilon$ explains the signal, but according to the definition of the full presence of photon in a particular place, we require an order 1 weak value of some local variable.  In view of this, we have only \lq secondary presence\rq\ \cite{pastsec} of the photon in $E$ in the present case. 

Now,  when mirror $E$ is tilted, we get $(O{\rm \bf P}_F)_w\simeq 2\epsilon$, indicating that the
presence of the photon is of order $\epsilon$ also at mirror $F$.   Nonetheless,  we do not get the peak at $f_F$ similar to that at $f_E$ by tilting mirror $F$ as well as $E$.
The reason for this is that the record of the interaction reaching the detector from the tilting mirror $F$ is of order $\epsilon$ and not of the order 1 as for the signal from mirror $E$.  Note that, when only mirror $F$ is tilted, we have $(O{\rm \bf P}_F)_w=0$. 

We have shown that the results of the interference experiment with a nested interferometer and a Dove prism inside it can be explained in the framework of the recently proposed approach \cite{past}. We get signals from mirrors $A$, $B$, and $C$, because the photon presence there is of order 1 and the trace recorded on the photon itself is of order $\epsilon$. A similar signal is obtained from mirror $E$ where the presence of the photon is of the order $\epsilon$, but recorded trace is of order 1.

 The signal in $E$ should disappear if the mirror will be wiggled in the perpendicular direction. If only this mirror is wiggled and everything else is not, then there will be exactly zero presence at $E$.
 If all mirrors are wiggled as in the experiment \cite{Danan}, then the presence will be of order of $\epsilon$, but the record will also be of order $\epsilon$, so  the signal will be too small to observe.
It will be of interest to perform a nested Mach-Zehnder interferometer experiment with wiggling mirrors and the Dove prism to demonstrate these effects.

\section{Bohmian trajectory}

\begin{figure}[t]
\begin{center}
\includegraphics[width=7cm]{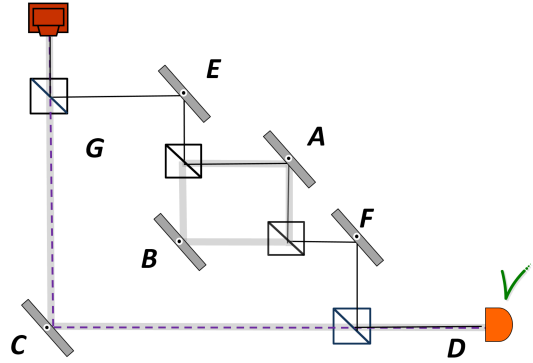}
\caption{Nested Mach-Zehnder interferometer tuned to destructive interference towards $F$ when a single photon is detected in $D$. Dashed line represents a common sense proposal by Wheeler, thick gray line describes the past according to Vaidman's proposal as places where the particle leaves a weak trace, continuous line represents the Bohmian trajectory.}
\label{fig:FIG-3}
\end{center}
\end{figure}

Before concluding let us analyse this nested interferometer in the framework of the Bohmian interpretation of quantum mechanics \cite{Bohm}.  While Bohr preached not to ask where  the particle inside the interferometer was, Wheeler suggested a \lq common sense\rq\, proposal based on classical intuition.
While we have suggested to rely on the weak trace that the particle leaves using the TSVF, Bohm has a proposal for a deterministic theory which associates a unique trajectory for every particle. In a particular case of nested interferometer which we consider, with or without Dove prism, the particle detected in $D$ has a well defined trajectory, see Fig.\ref{fig:FIG-3}. Note that it corrects an erroneous trajectory in Fig.2 of \cite{Berge60}. The simplest way to understand why Bohmian trajectory must be as shown, is the observation that Bohmian trajectories do not cross \cite{EnglertScully}. The probability to reach detector $D$ is only 1/9, while the probability to be in path $A$ is 1/3. Thus every Bohmian trajectory which reaches  $D$ had to pass through $A$.

Bohmian trajectories are entities beyond the standard quantum theory. One of us (L.V.) had a privilege to spend a day of discussions with David Bohm (Charlestone, SC, 1989).
I remember telling him what I liked in his theory: a consistent deterministic theory of everything, a candidate for a final theory. But he completely dismissed this approach. For him it was nonsense to look for a final theory.
He explained to me that his theory is just another step in infinite search for a better understanding of nature. He was certain that quantum theory is not the last word, and for finding a deeper and more precise theory, quantum theory has to be reformulated. His theory was a counter example to the wide spread belief generated by the von Neumann no-go theorem that it would be impossible to extend quantum mechanics consistently by adding hidden variables.
And, indeed, it opened new horizons for research.

\section{Conclusions}

Unless a quantum particle is not described by a well localized wave packet, the standard quantum theory cannot tell us where the particle was. Vaidman \cite{past} proposed the definition of where a quantum particle was according to the local trace it left: the particle was in a place where the trace is of the order of the trace a single localized particle would leave. In Danan {\it et al.} experiment photons told us  where they have been (according to the trace definition) in a specially tuned nested interferometer. The AJ modification of this experiment, {\it i.e.,} placing a Dove prism in one of the arms of the inner interferometer did not change significantly the past of the photons, but photons told a different story: they were also near mirror $E$ in spite of the fact that according to Vaidman's narrative they were not present there.  We conclude that the photons were lying about their presence in $E$,  in the sense that, although the trace they left there was much smaller than the trace that a localized photon would leave, the signal provided by the photons was large as if they had fully been present in $E$.

How could the photons produce the signal with frequency $f_E$ which was larger than any other signal? 
In the original and the modified experiments local traces were not observed. Instead,  locally created traces were \lq written\rq\ on the transversal degree of freedom of the photon itself. In the original experiment, the transversal degree of freedom was not distorted until it reached the detector, so these local traces were faithfully read by the detector. In the modified experiment, the Dove prism influenced the transversal degree of freedom spoiling the faithful readout of local traces by the detector. In fact, AJ mentioned such an interpretation in \cite{Jordan} as one of the options: \lq\lq One possible response to this result is that we have improperly read off the past of the photon by letting it suffer further interactions with the environment before reading the weak trace after it was written, so our weak measurement was a bad one for inferring the past of the photon.\rq\rq

Apart from the explanation of the experimental results  by the presence of the particle defined through the weak trace, Danan {\it et al.} presented a simpler argument of the presence of the photon in $A$, $B$ and $C$. The detected photons had to be there because they brought to the detector information which was only there. But the same should hold for the modified experiment: the particles had to be in $E$ because they brought information about $f_E$ which was present only in $E$. Sections IV and V explain how it happens in spite of the fact that the trace left by the particles at $E$ was very small. It was small, but not exactly zero as in the original experiment when only mirror $E$ was wiggling. The Dove prism did change the past of the photons a little. 

Introducing a Dove prism not only spoiled faithful transmission of the transverse degree of freedom of the photon to the detector, it also made the inner interferometer extremely sensitive for the misalignment of the input beam. The strength of the signal in the experiment was proportional to the relative amount of the orthogonal component created by local interaction. This component was the asymmetric mode which with the Dove prism passed in full through the inner interferometer, while the reference, the symmetric mode, did not pass at all due to the destructive interference. This explains how a small presence of the photons in $E$ caused a strong signal with frequency $f_E$ .

Note that the Bohmian trajectory did pass through $E$. But it also passed through $F$, although no frequency $f_F$ was observed. It is well known, starting from \lq surrealistic trajectories\rq\ \cite{EnglertScully}, that we cannot view quantum particles as acting locally in their Bohmian positions, see also \cite{Naaman}.

We have observed that introducing the Dove prism into inner interferometer of the Danan {\it et al.} experiment creates a tiny presence of the photons in $E$. 
However, we argue that from this we should not tell that the Dove prism changes the past of a photon in the nested interferometer proposed in \cite{past}.  
In fact, the origin of the presence of the photons can be found in the disturbance of the mirror $E$.  The weak value of any local operator at $E$ is strictly zero in an ideal interferometer
where no mirror is tilted, even if  the Dove prism is there.

\acknowledgments{This work has been supported in part by the Israel Science Foundation Grant No. 1311/14,
the German-Israeli Foundation for Scientific Research and Development Grant No. I-1275-303.14,
and the KEK Invitation Program of Foreign Scholars.}

\bibliographystyle{mdpi}

\renewcommand\bibname{References}


\end{document}